\documentclass[preprint,showpacs,preprintnumbers,amsmath,amssymb,endfloats*]{revtex4}

\usepackage{graphicx}

\begin{document}

\preprint{JAP}

\title{Superconducting properties of ultrathin $\mathbf{Bi_2Sr_2CaCu_2O_{8+}}{_x}$ single crystals}

\author{L. X. You}
\email{lixing@mc2.chalmers.se}
\author{A. Yurgens}
\author{D. Winkler}

\affiliation{Quantum Device Physics Laboratory, Department of
Microtechnology and Nanoscience, Chalmers University of Technology,
SE-412 96 G\"oteborg, Sweden}

\author{C. T. Lin}
\author{B. Liang}
\affiliation{Max-Planck-Institut f\"ur Festk\"orperforschung,
Heisenbergstrasse 1, D-70569 Stuttgart, Germany}

\date{\today}

\begin{abstract}
We use Ar-ion milling to thin $\mathrm{Bi_2Sr_2CaCu_2O_{8+}}{_x}$
(Bi2212) single crystals down to a few nanometers or one-to-two
$\mathrm{(CuO_2)_2}$ layers. With decreasing the thickness,
superconducting transition temperature gradually decreases to zero
and the in-plane resistivity increases to large values indicating
the existence of a superconductor-insulator transition in ultrathin
Bi2212 single crystals.
\end{abstract}

\pacs{74.25.Fy, 74.62.Yb, 74.72.Hs, 74.78.Bz, }

\maketitle

\section{Introduction}

Many high-temperature superconductors (HTS) are widely considered as
two-dimensional (2D). The cuprate HTS consists of conducting layers
of $\mathrm{CuO_2}$ planes separated by poorly conducting or even
insulating charge reservoirs. Among the most anisotropic HTS is
$\mathrm{Bi_2Sr_2CaCu_2O_{8+}}{_x}$ (Bi2212) compound, where the
separating charge reservoir consists of BiO$_2$ and CaO planes.
These build up a relatively large distance (12\AA) between
$\mathrm{(CuO_2)_2}$ planes which in turn results in a weak $c$-axis
coupling and strong anisotropy in both the normal and
superconducting state. This high anisotropy and the
quasi-two-dimensional (2D) character of conductivity and
superconductivity of $\mathrm{CuO_2}$ planes result in a number of
unusual physical properties and effects, with intrinsic Josephson
tunneling standing out as one of the most intriguing one
\cite{Kleiner:PRL92}.

The Kosterlitz-Thouless (KT) transition which appears substantially
below the bulk transition temperature $T_{c0}$ and is associated
with the thermal dissociation of vortex-antivortex pairs above a
certain temperature $T_\mathrm{KT}$ is characteristic for a
disorder-free 2D superconductor. As the disorder is introduced,
$T_\mathrm{KT}$ can be further suppressed \cite{Fisher:PRL90}, even
down to zero at a certain critical disorder strength or below a
specific thin-film thickness \cite{Haviland:PRL89}. Several
publications have shown that thin films of different superconducting
materials become insulating when their normal-state resistance is
larger than the universal quantum resistance $R_q=h/(2e)^2\approx
6.5\ {\mathrm k\Omega}$ \cite{Cha:PRB91, Jaeger:PRB86,
Ferrell:PRB88}, i.e. a superconductor-insulator transition occurs.

The question naturally arises, whether the superconductivity of an
isolated $\mathrm{(CuO_2)_2}$ plane is sufficiently robust and is
characterized by a non-zero $T_{\mathrm KT}$. To provide
experimental evidence for this issue is obviously a very difficult
and challenging task. The epitaxial cuprate film layers interleaved
between buffering thin-film layers has been one way of investigating
the superconducting properties of isolated $\mathrm{(CuO_2)_2}$
planes \cite{Ota:PhysicaC99, Qi:JLTP99, Boze:JCG00, Bozovic:ASC01}.
However, the initial growth of a film involves complicated
nucleation processes affected by the lattice mismatch between the
film and substrate~\cite{Ohring:Book}. For a better lattice matching
one can grow buffer layers of non-superconducting Bi2201 between
Bi2212 and the substrate~\cite{Ota:PhysicaC99}. This buffer contains
$\mathrm{CuO_2}$ planes and electrically is quite well conducting,
which may affect the superconductivity in the ultrathin Bi2212 films
on top.

HTS bulk single crystals with perfect crystal structure are
routinely grown by many groups \cite{Hardy:PhysicaC94,
Pissas:SUST97, Lin:PhysicaC00, Yao:SUST04}. Bi2212 single crystals
are widely used in many studies due to their high anisotropy and the
presence of the intrinsic Josephson effect~\cite{Kleiner:PRL92}. The
thickness of the single crystals is always very large compared with
thin films and is difficult to measure or control precisely.
However, the indisputable merit of single crystals is that the
crystal orientation is perfect in all three dimensions and there are
no complications with regard to the nucleation processes or any
lattice mismatch characteristic between the thin films and the
substrate.

In this work, by using conventional photolithography and Ar-ion
milling we could controllably thin down single crystals to any
thickness. We see that the superconducting critical temperature
$T_c$ does not depend on thickness down to a few nanometers. The
superconductivity gradually vanishes on further decrease of the
thickness.

\section{Sample preparation}

Bi2212 single crystals with a typical critical temperature $T_c \sim
85 $ K were grown using the traveling solvent floating zone (TSFZ)
method \cite{Lin:PhysicaC00}. The main fabrication process is
similar to the process for  fabrication of IJJs'~\cite{Wang:IEICE02,
You:JJAP04,You:SUST04}.

\par
First, we glue a single crystal of Bi2212 onto a sapphire substrate
using polyimide. Then we cleave the crystal using common Scotch
tape. Immediately after the cleavage, the single crystal is covered
by a 20 nm thin film of $\mathrm{CaF_2}$ followed by 20 nm of gold.
Both films are made by physical-vapor deposition in the same chamber
without breaking vacuum. $\mathrm{CaF_2}$ with strong ionic bonds
evaporates as a molecule and appears to be chemically inert to
HTS~\cite{Ginsberg:book}.  The Au thin-film is needed to provide
higher optical contrast when the single crystal becomes so thin that
it is almost 100 \% transparent for visible light while the
intermediate $\mathrm{CaF_2}$ layer is needed to protect and isolate
the Bi2212 surface from the Au layer.

\begin{figure}
\includegraphics[width= 8.5 cm]{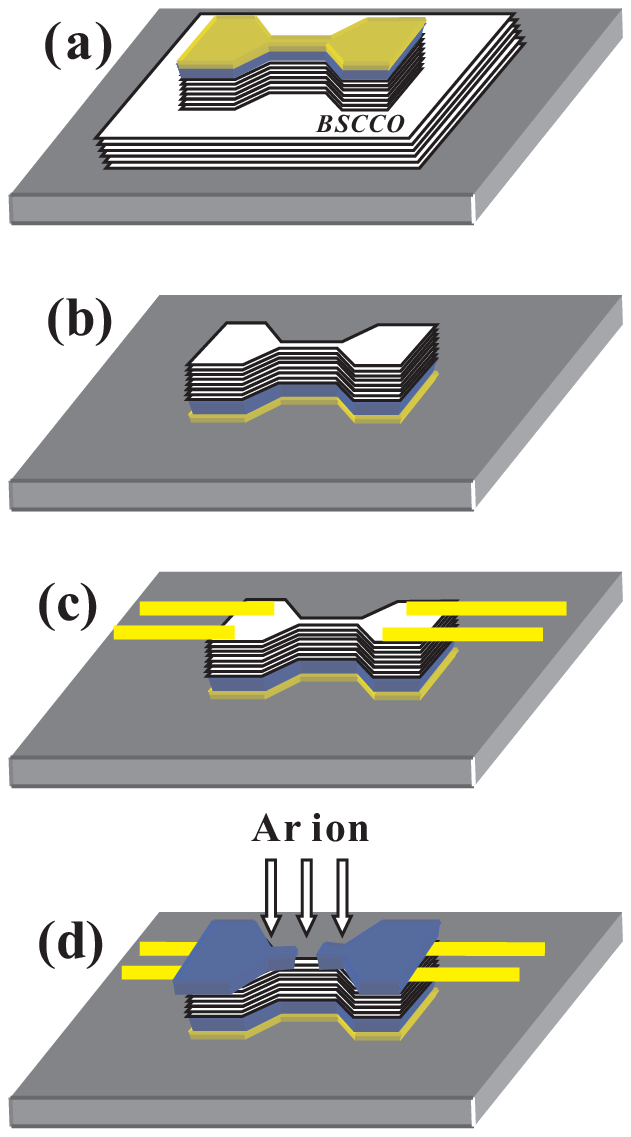}
\caption{(Color online) Schematic view of the patterning steps for
the micro-crystal and a bridge in the center.} \label{process}
\end{figure}

By conventional photolithography and Ar-ion etching, a bow-tie
shaped mesa with a micro-bridge in the center is formed on the
crystal (see Fig.~\ref{process}(a)). The overall thickness of the
Bi2212 mesa is typically about 100~nm which is controlled by the
etching time and rate. In the next step, we flip the sample and glue
it by using polyimide to another sapphire substrate, sandwiching the
single crystal between the two substrates. Separating the substrates
cleaves the single crystal into two pieces. One piece has the mesa
at the bottom which is now upside down and faces the substrate. We
remove all material but the mesa by iteratively cleaving the former
with the aid of Scotch tape and inspecting the resulting sample in
an optical microscope. The schematic view of the resulting sample
where the stand-alone ``ex"-mesa is only left is shown in Fig.
\ref{process}(b).

Another Au layer is then deposited and patterned immediately after
that to make four gold electrodes to this tiny piece of the single
crystal (micro crystal)(see Fig. \ref{process}(c)). Usually we
slightly ``over" etch this Au layer to make sure that no residue of
gold is left on the surface between the electrodes. The micro-bridge
and other areas of the micro crystal outside the electrodes
therefore get slightly thinner. The micro-bridge is further thinned
down by the subsequent Ar-ion etching, while electrodes and areas in
between are usually protected from etching by an additional
patterned $\mathrm{CaF_2}$ layer (see Fig. \ref{process}(d))

Using these contacts, we could continuously measure the resistance
of the bridge \textit{in situ}, during the etching at room
temperature. The superconducting properties of the sample were
measured after each etching in a separate cryogenic system.

\section{Sample topography}

\begin{figure}
\includegraphics[width= 8.5 cm]{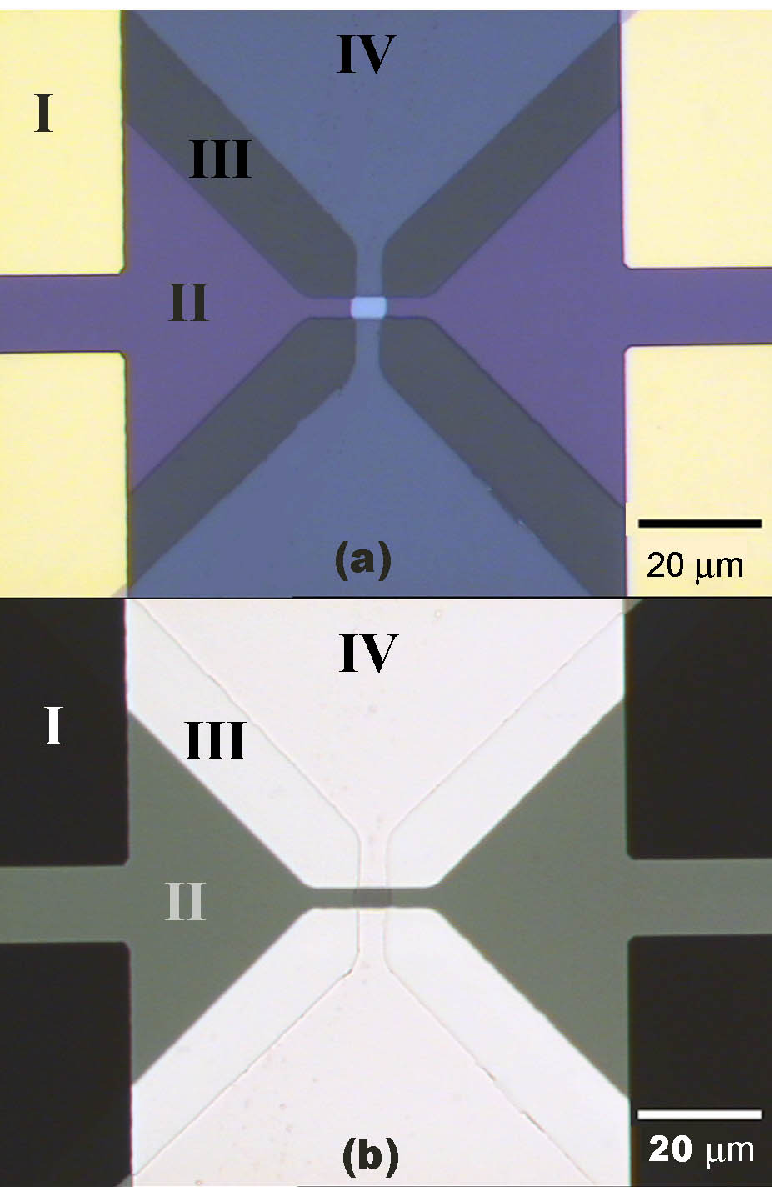}
\caption{(Color online) (a), (b) Optical images of a sample
illuminated from the top and bottom respectively. Indicated in the
images are I) the gold contact, II) the Bi2212 micro-bridge, III)
the $\mathrm{CaF_2}$ layer and IV) the substrate. All the gold
contacts and the micro-bridge are protected by $\mathrm{CaF_2}$,
except for the rectangle ($\mathrm{7 \times 3.5~\mu m^2}$) in the
middle of the bridge.} \label{optics}
\end{figure}

Fig. \ref{optics} shows optical images of a sample illuminated from
the top (a) and bottom (b). The width of the micro-bridge is 3.5
microns and the open area at the bridge for further thinning is 7
microns long. The contrast of the images is high owing to the gold
thin film in the bottom of the structure.

\begin{figure}
\includegraphics{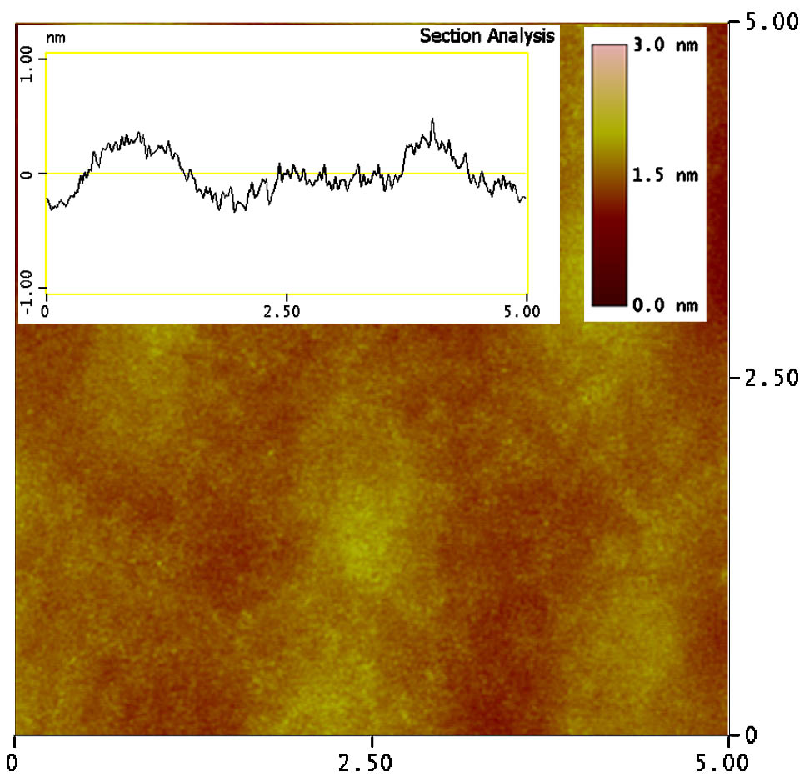}
\caption{(Color online) The AFM image over an area of $\mathrm{5
\times 5\ \mu m^2}$ on a Bi2212 micro-bridge. The thickness of the
bridge is about 30 nm. The inset shows the surface profile along an
arbitrarily chosen direction.} \label{AFM}
\end{figure}

It is quite important to assure that the Ar-ion etching is uniform.
Atomic force microscopy (AFM) was used to examine the surface
topography of one of the fabricated 30 nm thick micro-bridges. Fig.
\ref{AFM} shows an AFM image across a $\mathrm{5\times 5~\mu m^2}$
large area of a wider micro-bridge. The surface of the bridge is
quite flat with an rms roughness of 0.28 nm and a mean surface
roughness of 0.21 nm. This roughness is quite close to the freshly
cleaved surface of a Bi2212 single crystal with an rms surface
roughness of 0.20 nm~\cite{You:PRB05}. For Bi2212 thin films with a
similar thickness prepared by MBE, the mean surface roughness is
between 0.5 and 0.9 nm across an area of $\mathrm{10\times 10~\mu
m^2}$ \cite{Bove:JCG00}. It is clear that even after the Ar-ion
etching the surface quality of the single crystals  is better than
that of thin films of comparable thickness grown by MBE. The surface
quality of several samples of different thicknesses was examined,
all with about the same result.

\section{Electronic measurements and discussions}

\begin{figure}
\includegraphics[width= 8.5 cm]{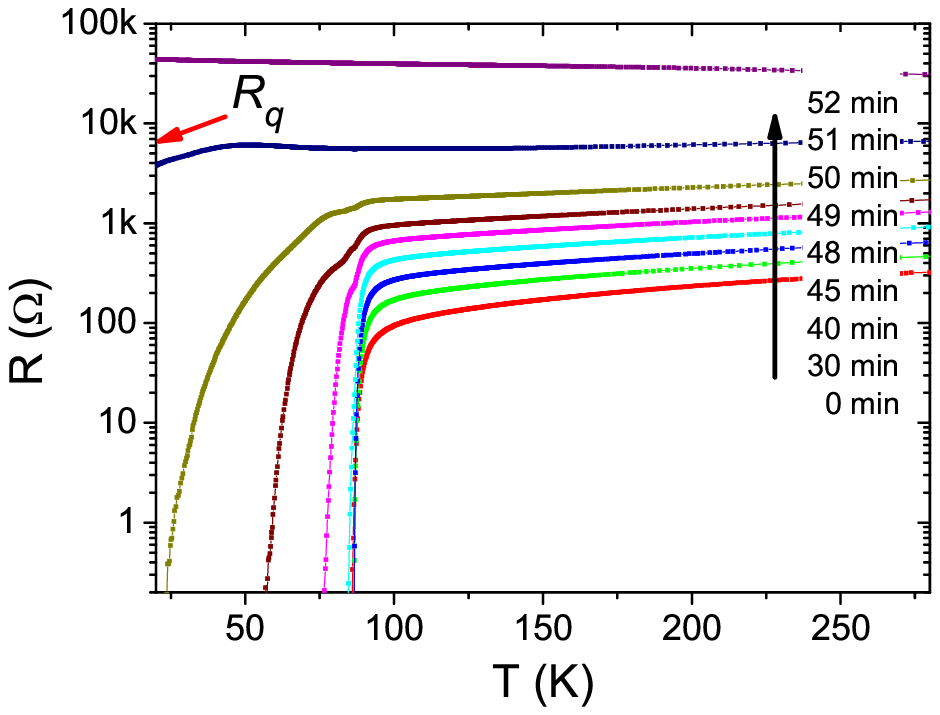}
\caption{(Color online) RT curves of a BSCCO micro-bridge with the
open area $\mathrm{3.5 \times 3.5\ \mu m^2}$ after subsequent
etching times. The initial thickness of the bridge is less than 90
nm.} \label{RT}
\end{figure}

The electrical transport measurements of the micro-bridges were
carried out in a temperature range of $16-290$~K. Fig. \ref{RT}
shows the $R-T$ curves of a micro-bridge with a $\mathrm{3.5 \times
3.5~\mu m^2}$ large unmasked area in the middle of the bridge after
each of the subsequent etchings. Despite the vanishingly small
thickness, the micro-bridges are very stable and withstand more than
ten cycles of cooling down from room temperature to 16 K without any
noticeable change in their resistance.

The typical etching parameters used in the experiment are 230 V for
the Ar-ion acceleration voltage and 0.11 mA/cm$^2$ ($\mathrm{7\times
10^{14} ~s^{-1}cm^{-2}}$) for the beam current density, resulting in
an etching rate $\alpha$ of about 1.5 nm/min or about half a unit
cell in $c$-axis per minute.

When the total etching time is less than 40 minutes, the
superconducting critical temperature $T_{c0}$ is the same as for a
bulk single crystal ($\sim$86 K). After some additional etching for
about 8 minutes, $T_{c0}$ starts to decrease. $T_{c0}$ rapidly
decreases to 25~K after just two more minutes of etching and finally
the specimen ceases to be superconductive. At the same time, the
temperature dependence of the resistance changes from a metal- to a
more semiconductor-like behavior above $T_c$, and shows the presence
of a Superconductor-Insulator (S-I) transition below $T_c$.

The S-I transition is an important issue in condensed matter physics
\cite{Imada:RMP98}. For an ultrathin film, the S-I transition occurs
when the sheet resistance $R_\Box$ is about or larger than the
universal quantum value $R_q=h/4e^2\approx 6.5~\mathrm{k\Omega}$
\cite{Jaeger:PRB86, Ferrell:PRB88}.

The most resistive part of the bridge is the etched (and thinnest)
square area $\mathrm{3.5 \times 3.5~\mu m^2}$ in the middle of the
bridge, and the resistance of the other parts in the bridge is much
smaller and negligible compared with the resistance in the middle
part. As a result, the resistance which we measure is close to the
sheet resistance $R_\Box(d)$ for the thinnest bridge. The quantum
resistance is indicated by an arrow in Fig. \ref{RT}, and we see
that the boundary value for S-I transition in our samples is
consistent with $R_q$. Similar S-I transition was also observed in
other HTS ultrathin films
\cite{Qi:JLTP99,Kabasawa:JAP96,Salluzzo:PRB04}.

\par
Given that the etching rate is constant, we can calculate the
thickness of the bridge from the etching time provided we know an
initial thickness of the bridge. Although we could roughly measure
the initial thickness by, say, AFM, the absolute error would be too
large for a self-consistent analysis of the resistance measurements
for different thicknesses. The polyimide layer which is used to glue
the micro-bridge to the substrate is not flat and sufficiently
smooth to be used as the reference plane in the thickness
measurements of the micro-bridge.

Nonetheless, we can use the whole set of resistance-vs-etching time
data to deduce the unknown initial thickness assuming the uniform
in-plane resistivity $\rho_{ab}$. The total resistance of the
micro-bridge $R$ consists of two parts, the in-plane resistance
$R_t$ of the middle thinner part and the resistance $R_s$ of the two
surrounding thicker parts. Strictly speaking, we should take into
account some contribution from the $c$-axis resistivity also because
the electrodes are only attached to one side of the highly
anisotropic single crystal and the bias current should flow along
the $c$-axis before getting into the bridge~\cite{Busch:PRL92}.
However, since the area of the electrodes is relatively large, we
can ignore this contribution and assume it is just a small part of
$R_s$.

Let the thinner middle part of the bridge be a slab with the width
$w$, thickness $d$ and length $l$. Its resistance $R_t=\rho_{ab}
l/wd=\rho_{ab}/d$ for $l=w$ (square).  The thickness is assumed to
be a linear function of the etching time $t$: $d=d_0-\alpha t=\alpha
(t_0-t)$, where $d_0$ is the initial thickness and $t_0$ is the
total time needed to etch clear through it.  The total resistance is
$R(t)=R_s+\rho_{ab}/\alpha (t_0-t)$.

\begin{figure}
\includegraphics[width= 8.5 cm]{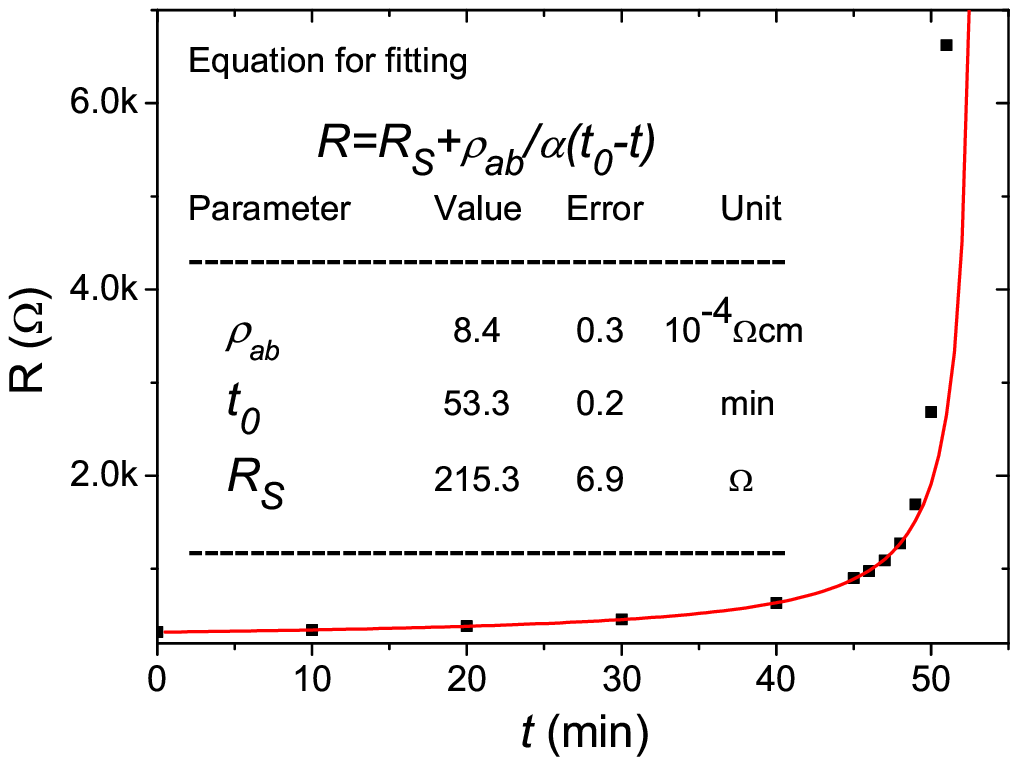}
\caption{Resistance (T = 273 K) of the micro-bridge vs. the etching
time (dots) fitted by $R(t)=R_s+\rho_{ab}/\alpha(t_0-t)$ (solid
line). Best-fit parameters are shown in the inset. $\alpha=1.5$
nm/min was measured in seperate experiments.} \label{Rfit}
\end{figure}

This equation is used to fit the experimental $R(t)$ from
Fig.\ref{RT} using $R_s$, $\rho_{ab}$, and $t_0$ as fitting
parameters. $\alpha=1.5$~nm/min was accurately measured in a
separate experiment. As is seen in Fig.\ref{Rfit}, the fit is quite
good for etching times up to 48 min, while it becomes worse for
times longer than that. This can be due both to the close proximity
to the S-I transition and possibly to deteriorated properties of the
last $\mathrm{(CuO_2)_2}$ layer. In the latter case, the last
$\mathrm{(CuO_2)_2}$ layer had once been the surface layer before
the $\mathrm{CaF_2}$ layer was deposited on it, see above.
Inevitable exposure to moisture in the air can result in worsening
of the surface layer~\cite{Kim:PRB99,Zhu:PhysicaC}.

From previous experience and present measurements, we know that
there is no more than {\it one} surface layer which is usually
affected by moisture or by the contact to a normal metal. This can
easily be verified from measuring the $I-V$ curves of a stack of IJJ
in a three-probe measurement at low temperatures, see for instance
Fig. 2a in Ref.~\cite{Kim:PRB99}. Only the first branch which
corresponds to the surface junction has a reduced critical current.
The adjacent junction already has the nominal critical current equal
to the critical current of the majority of junctions in the stack.

From the best least-squares fit, $\rho_{ab}$, $R_s$, and $t_0$ could
be determined, see Fig. \ref{Rfit}. The in-plane resistivity
$\rho_{ab}$ at 273 K is about $8\times 10^{-4}\ \mathrm{\Omega \cdot
cm}$, which is close to $\rho_{ab}$ reported
elsewhere~\cite{Cooper:Nature90,Watanabe:PRL97}. Knowledge of $t_0$
and $\alpha$ allows us to determine the precise thickness
corresponding to each etching time. The initial thickness can be
also calculated to be about 80~nm, which is not far from the
estimations based on the etching time in the first step of the mesa
fabrication, see Fig.~1a above.

\begin{figure}
\includegraphics[width= 8.5 cm]{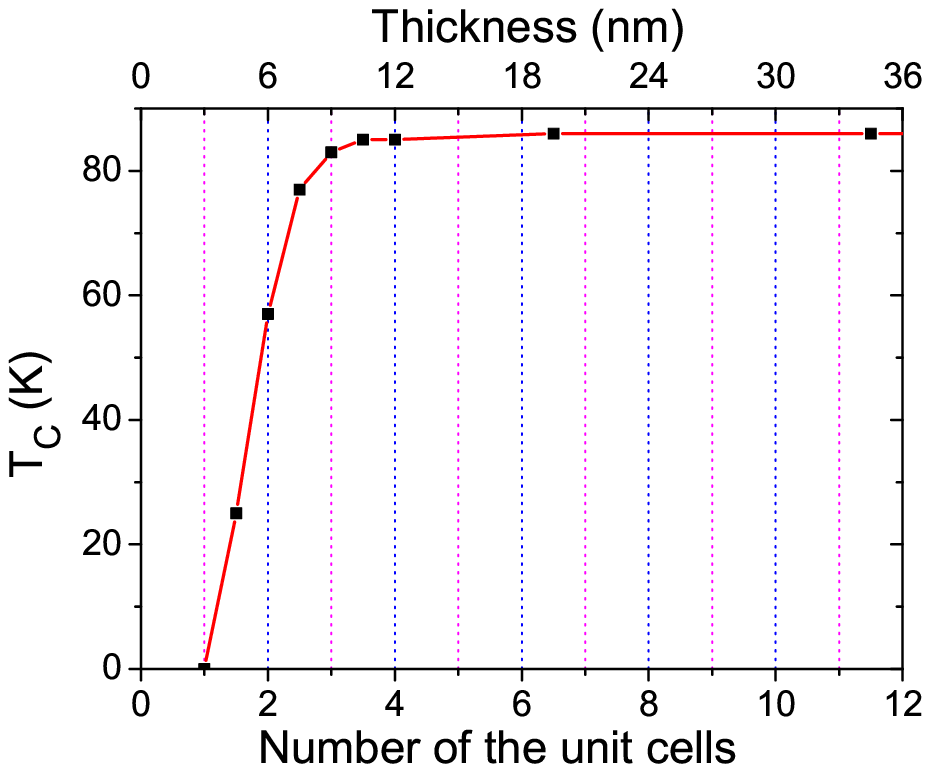}
\caption{Variation of $T_{c0}$ with the thickness of the bridge
(number of unit cells) in $c$-axis direction.} \label{TcN}
\end{figure}

Fig. \ref{TcN} shows the variation of $T_{c0}$ with the thickness
expressed both in nm and the number of unit cells. When the
ultrathin Bi2212 single crystal has more than 8
$\mathrm{(CuO_2)_2}$-layers (12 nm or 4 unit cells thick), it has
the same $T_c$ as the bulk. $T_c(d)$ starts to decrease when it is
thinner than 4 unit cells, but superconductivity with finite
$\mathrm{T_{c0}}=25$~K still persists even when there are just 1.5
unit cells left ($\sim$ 5 nm). The suggested S-I transition occurs
in a slightly thinner bridge (3 nm). We have to mention here that
the thickness we obtained from the curve fitting is an effective
value. The physical thickness should include the thickness of a
surface insulating layer formed in the ion etching process, which is
no more than 3~nm \cite{You:SUST03}.

Our result that the superconductivity of ultrathin Bi2212 single
crystals vanishes suggests that these are intrinsically disordered.
This is in agreement with several STM studies (Ref.
\onlinecite{Lang:Nature02}, for instance) revealing short-range
disorder in Bi2212 single crystals cleaved at cryogenic
temperatures, as well as with other experiments on Bi2212 thin films
demonstrating S-I transition \cite{Ota:PhysicaC99, Qi:JLTP99}.
However, experiments with ultrathin $\mathrm{YBa_2Cu_3O_{7+\delta}}$
(YBCO) films sandwiched between PrBCO buffer layers showed that a
one-unit-cell-thick YBCO film had a non-zero $T_\mathrm{KT}\approx
30$~K \cite{Matsuda:PRB93} . This can be due to better quality of
RHEED-controlled YBCO epitaxial films showing less disorder but
might also be due to the presence of relatively thick and
electrically conducting PrBCO buffer- and cap layers.

\section{Conclusions}

With conventional micro-fabrication processing high-quality
ultrathin Bi2212 single crystals were fabricated and studied.
Superconductivity was observed for all thicknesses down to
effectively 1.5 unit cells. In the thinner crystals, the
superconductivity quenches while $R(T)$ changes from a metallic to a
semiconductor-like behavior, suggesting a superconductor-insulator
transition that takes place around $R_\Box \sim R_q=6.5\
\mathrm{k\Omega}$.

\begin{acknowledgments}

We thank M. Torstensson and D. Lindberg for technical assistances
and fruitful discussions. This work is financed by The Swedish
Foundation for Strategic Research (SSF) through the OXIDE program.

\end{acknowledgments}

\newpage

\section*{Figure Captions}

Figure \ref{process}. (Color online) Schematic view of the
patterning steps for the micro-crystal and a bridge in the center.

Figure \ref{optics}. (Color online) (a), (b) Optical images of a
sample illuminated from the top and bottom respectively. Indicated
in the images are I) the gold contact, II) the Bi2212 micro-bridge,
III) the $\mathrm{CaF_2}$ layer and IV) the substrate. All the gold
contacts and the micro-bridge are protected by $\mathrm{CaF_2}$,
except for the rectangle ($\mathrm{7 \times 3.5~\mu m^2}$) in the
middle of the bridge.

Figure \ref{AFM}. (Color online) The AFM image over an area of
$\mathrm{5 \times 5\ \mu m^2}$ on a Bi2212 micro-bridge. The
thickness of the bridge is about 30 nm. The inset shows the surface
profile along an arbitrarily chosen direction.

Figure \ref{RT}. (Color online) RT curves of a BSCCO micro-bridge
with the open area $\mathrm{3.5 \times 3.5\ \mu m^2}$ after
subsequent etching times. The initial thickness of the bridge is
less than 90 nm.

Figure \ref{Rfit}. Resistance (T = 273 K) of the micro-bridge vs.
the etching time (dots) fitted by $R(t)=R_s+\rho_{ab}/\alpha(t_0-t)$
(solid line). Best-fit parameters are shown in the inset.
$\alpha=1.5$ nm/min was measured in seperate experiments.

Figure \ref{TcN}. Variation of $T_{c0}$ with the thickness of the
bridge (number of unit cells) in $c$-axis direction.

\end{document}